\def\b{\bibitem}
\def\boldphi{\mbox{\boldmath $\phi$}}
\def\boldvarphi{\mbox{\boldmath $\varphi$}}
\documentstyle[aps,prb,eqsecnum,epsf,floats]{revtex}

\begin{document}

\twocolumn[\hsize\textwidth\columnwidth\hsize\csname@twocolumnfalse%
\endcsname
\title{The effect of rare regions on a disordered itinerant
       quantum antiferromagnet with cubic anisotropy}
\author{Rajesh Narayanan}
\address{Theoretical Physics, University of Oxford, Oxford, OX3 NP1, UK}
\author{Thomas Vojta}
\address{Institut f{\"u}r Physik, TU Chemnitz, D-09107 Chemnitz, Germany}
\date{\today}
\maketitle
\begin{abstract}
We study the quantum phase transition of an itinerant antiferromagnet
with cubic anisotropy in the presence of quenched disorder,
paying particular attention to the locally ordered spatial regions
that form in the Griffiths region. We derive an effective action where
these rare regions are described in terms of static annealed disorder.
A one loop renormalization group analysis of the effective action shows that
for order parameter dimensions $p<4$ the rare regions destroy the
conventional critical behavior. For order parameter dimensions $p>4$
the critical behavior is not influenced by the rare regions, it is described
by the conventional dirty cubic fixed point. We also discuss the influence
of the rare regions on the fluctuation-driven first-order transition in
this system.
\end{abstract}
]
\section{Introduction}
\label{sec:I}
Quenched disorder can have very drastic influences on the critical behavior
of a system undergoing a continuous phase transition.
According to the Harris criterion \cite{Harris} the critical behavior
of a clean system is unaltered by disorder, if the correlation length
critical exponent $\nu$ obeys the inequality $\nu > 2/d$, where $d$
is the spatial dimensionality of the system.
In the opposite case, $\nu < 2/d$, the clean critical behavior is unstable,
and the disorder either leads to a new, different universality class, or to
an unconventional critical point, or even to the destruction of the phase
transition.

Another, less well understood consequence of quenched disorder is the
formation of rare locally ordered regions in the disordered phase.
For a transition occuring at a finite temperature,
this can be explained in the following way.
In general, disorder leads to the suppression of the critical temperature
from its clean value $T_c^0$ to $T_c$. In the temperature region
between $T_c^0$ and $T_c$ the system does not show long-range order.
However, there will be arbitrarily large regions which are devoid of
impurities and thus order locally. The probability of finding such regions
usually decreases exponentially with their size, they represent
non-perturbative degrees of freedom. These locally ordered regions
are known as rare
regions, and the order parameter fluctuations induced by them as
local moments or instantons.
Griffiths \cite{Griffiths} showed that the rare regions
lead to a non-analytic free energy everywhere in the
temperature region between $T_c^0$ and $T_c$, now called
the Griffiths region or Griffiths phase.
In generic classical systems this is a very weak
effect, and the non-analyticity in the free energy is only an
essential one.
However, the Griffiths singularities become stronger if the
disorder is spatially correlated. McCoy and
Wu \cite{McCoy} studied a two-dimensional Ising model where
the disorder is perfectly correlated in one spatial
direction and uncorrelated in the other. In this model
the rare regions lead to the divergence of the susceptibility
at some temperature $T_{\chi}$ within the Griffiths region.

A very interesting question is what is the influence of the
rare regions on the critical behavior of a system.
Dotsenko et al. \cite{Dotsenko} studied this question for
a weakly disordered classical ferromagnet. They found
that the conventional theory of critical behavior \cite{Grinstein}
in this system is unstable with respect to replica symmetry breaking.
They also showed that the rare regions actually induce replica symmetry
breaking perturbations and thus destabilize the conventional
critical fixed point. While so far no final conclusion about
the fate of the transition in the weakly disordered ferromagnet
could be reached, the occurrence of replica symmetry breaking
raises the possibility of an unconventional transition with
activated scaling, as is believed to occur in the random field
Ising model \cite{random-field}.

For quantum phase transitions \cite{QPT} which occur at zero temperature
as a function of some non-thermal control parameter, one expects an even
stronger influence of the rare regions than for classical transitions.
The reason is that a quantum model with uncorrelated quenched disorder
is effectively equivalent to a classical model with the disorder being
perfectly correlated in one dimension (the imaginary time dimension).
Fisher \cite{Fisher} investigated the critical behavior
of a one-dimensional quantum Ising spin chain in a
transverse field which is equivalent to the classical McCoy-Wu model.
He found that due to the rare regions the critical behavior is
of the activated form. This has been confirmed by numerical simulations
\cite{num1D} which also suggest \cite{num2D} that this
sort of behavior may not be restricted to one-dimensional systems.

In two recent papers \cite{us_rr} we have considered the effect of rare regions
on quantum phase transitions of itinerant electrons in $d>1$.
We have developed a systematic approach, representing the local moments by
inhomogeneous saddle point solutions of the field theory. The interaction
between the local moments and the fluctuations leads to a new term
in the effective action which is of the form of annealed static disorder.
In the case of the quantum antiferromagnetic transition
this new term results in the destruction of the
conventional critical fixed point if the number $p$ of order
parameter components is smaller than 4. No new fixed point
could be identified, the system displays runaway flow to
large disorder strength.
On the other hand, for the quantum ferromagnetic transition
the rare regions do not affect the critical behavior since
a self-induced long-range interaction suppresses all fluctuations
including those produced by the local moments.

In this paper we apply the approach developed in Ref. \onlinecite{us_rr}
to a model of an itinerant antiferromagnet with an additional interaction term
with cubic symmetry.
This model is equivalent to a weakly disordered classical ferromagnet
with cubic anisotropy in which the disorder is perfectly correlated
in some of the spatial dimensions but uncorrelated in the remaining dimensions.
The conventional theory for this model (without taking rare regions into
account) has been developed by Yamazaki, Holz, Ochiai and Fukuda\cite{Yamazaki}.

The purpose for this work is threefold.
We want investigate (i) whether the conventional critical fixed point
is stable under the influence of the rare regions. If it is unstable
we want to find out (ii) whether a new stable fixed fixed point exists
which describes a rare region driven transition. Finally we want to
study (iii) the influence of the rare regions on the fluctuation-driven
first-order transition occurring in our system.
The layout  of the paper is as follows. In
Sec.\ \ref{sec:II} we derive the effective field theory
by taking into account the disorder induced rare regions. In
Sec.\ \ref{sec:III}, we carry out the renormalization group analysis.
Finally, Sec.\ \ref{sec:IV} is left for a summary of our results.
\section{An effective action for disordered antiferromagnets with cubic anisotropy}
\label{sec:II}
\subsection{The model}
\label{subsec:II.A}

In 1976 Hertz \cite{Hertz} derived an order parameter field theory for the
description of the antiferromagnetic quantum phase transition of itinerant
electrons. Later this model was generalized to the dirty case by making the
distance from the critical point a random function of position
\cite{us_rr,bk_afm}. Here we consider an extension of this order
parameter field theory by incorporating an additional $\phi^4$ term which
possesses a (hyper-)cubic symmetry.

In terms of the $p$-component order parameter field $\boldphi$
(with components $\phi_i$) the total action can be written as
\begin{mathletters}
\label{eqs:2.1}
\begin{equation}
S[\boldphi ] = S_{\rm G}[\boldphi ]
 + S_{{\rm int}}[\boldphi]
+ S_{{\rm cubic}}[\boldphi]\quad,
\label{eq:2.1a}
\end{equation}
with the Gaussian part, the interaction part and the
cubic anisotropic part given by
\begin{equation}
S_{\rm G}[\boldphi ] = \frac{1}{2} \int dx\,dy\
  \sum_i \phi_i (x)\,\Gamma(x-y)\,\phi_i (y)\quad,
\label{eq:2.1b}
\end{equation}
\begin{equation}
 S_{{\rm int}} [\boldphi ] = u\int dx\,\sum_{i,j}
   \phi_i (x)\, \phi_i (x)\,
\phi_j (x)\, \phi_j (x)\quad,
\label{eq:2.1c}
\end{equation}
\begin{equation}
 S_{{\rm cubic}} [\boldphi ] = \lambda\, \int dx\,\sum_{i}
   {\phi_i}^4 (x)\, \quad.
\label{eq:2.1d}
\end{equation}
\end{mathletters}%
Here we use a 4-vector notation to combine the real space
coordinate ${\bf x}$ and imaginary time $\tau$, $x=({\bf x},\tau$),
$\int dx = \int d{\bf x} \int_0^{1/T} d\tau$. The bare
two point function,
\begin{eqnarray}
\Gamma({\bf x}-{\bf y},\tau-\tau')&=&\Gamma_0({\bf x}-{\bf y},\tau - \tau')
\nonumber\\
  &&+ \delta({\bf x}-{\bf y})\,\delta(\tau - \tau')\,\delta t({\bf x})\quad,
\label{eq:2.2}
\end{eqnarray}
consists of the deterministic part derived by Hertz \cite{Hertz}
whose Fourier transform reads
\begin{equation}
\Gamma_0({\bf q},\omega_n) = t_0 + {\bf q}^2 + \vert\omega_n\vert\quad,
\label{eq:2.3}
\end{equation}
and a disorder part in the form of a "random mass" term.
Here ${\bf q}$ is the wave vector, $\omega_n$ is a bosonic Matsubara
frequency and $\delta t({\bf x})$ is a random
function of position and is endowed with the following
statistical properties:
\begin{mathletters}
\label{eqs:2.4}
\begin{equation}
\langle\delta t({\bf x})\rangle = 0\quad,
\label{eq:2.4a}
\end{equation}
\begin{equation}
\langle\delta t({\bf x})\,\delta t({\bf y})\rangle = \Delta\,\delta({\bf x}
                                                        - {\bf y})\quad.
\label{eq:2.4b}
\end{equation}
\end{mathletters}%

\subsection{Inhomogeneous saddle points and annealed disorder}
\label{subsec:II.B}

In the conventional approach to critical behavior in systems
with quenched disorder \cite{Grinstein} the disorder average
is carried out at the beginning of the calculation by means of the
replica trick \cite{replica}. A subsequent perturbative analysis of  the
resulting, spatially homogeneous effective theory misses the
rare regions we are interested in since they are non-perturbative
degrees of freedom.

We therefore follow the approach developed in Ref. \onlinecite{us_rr},
and work with a particular realization of the disorder rather than
integrating it out. Let us consider spatially inhomogeneous, but
time-independent saddle point solutions of the action (\ref{eqs:2.1})
(time-dependent saddle-point solutions -- if any -- will always have a higher
free energy since the disorder is static).
Depending on the sign of the cubic interaction term the structure
of the saddle points in the $p$-dimensional order parameter space will be
different. When $\lambda>0$ the
free energy is minimized by saddle point solutions that
lie on the diagonals of a $p$-dimensional hypercube, while
when $\lambda<0$ the free energy is minimized by solutions
that lie on the axis of the hypercube.
In either case the modulus $\phi_{\rm sp}$ of these minimizing
saddle point solutions fulfills the equation
\begin{mathletters}%
\label{eqs:2.5}
\begin{equation}
\left(t_0 + \delta t(\bf x) - \partial_{\bf x}^2\right)\,|\boldphi_{\rm sp}
   ({\bf x})| + 4\,u_{\rm eff}\,|\boldphi_{\rm sp}({\bf x})|^3 = 0\quad,
\label{eq:2.5a}
\end{equation}
\begin{equation}
u_{\rm eff} = \cases{u + \frac{\lambda}{p} & for $\lambda>0$\cr
              u+\lambda & for $\lambda<0$\cr}\quad.
\label{eq:2.5b}
\end{equation}
\end{mathletters}%
Although $\phi_{\rm sp}({\bf x})=0$ is always a solution, there will
be spatially inhomogeneous solutions if $\delta t({\bf x})$ has
sufficiently deep and wide troughs \cite{us_rr}.
Let us now consider the Griffiths region, i.e. the region where
the average distance $t_0$ from the critical point is  positive
but where there are isolated islands  which support a non-zero
$\phi_{\rm sp}$.
If we have $N$ such islands which are sufficiently apart from each other
the global saddle point solutions may be written as
\begin{eqnarray}
\boldphi^{\{\sigma_I\}}_{\rm sp}({\bf x}) \equiv \Phi^{\{\sigma_I\}}({\bf x})
  &=&\sum_{I=1}^{N} \psi_I({\bf x})\,\sigma_I
\label{eq:2.7}
\end{eqnarray}
where $\psi_I({\bf x})$ is a solution of (\ref{eqs:2.5}) on the island $I$
and $\sigma_I$ is a unit vector in spin space (on one of the axis for
$\lambda<0$ or on one of the diagonals for $\lambda>0$).

Since the direction of the order parameter on each of the $N$ islands
can be chosen independently, (\ref{eq:2.7}) describes an exponentially
large number of degenerate saddle points, $(2p)^N$ for $\lambda<0$ and
$(2^p)^N$ for $\lambda>0$. To be precise, the saddle points are not exactly
degenerate due to the residual interaction of the (exponentially small) tails
of the order parameter between the islands.
The complicated structure of the free energy
landscape connected with the existence of an exponentially large number of
almost degenerate saddle points will finally turn out to be  responsible for
the failure of the conventional approach.

We now consider fluctuations around the saddle points (\ref{eq:2.7}).
Since the saddle points are separated by large free energy barriers
an expansion around one of them will not give a good representation
of the partition function of the entire system.
Instead we will restrict ourselves to small fluctuations and simply
add the contributions coming from {\em all} of the saddle points.
Thus the partition function for a particular realization $\delta t({\bf x})$
of the disorder can be written as
\begin{equation}
Z[\delta t({\bf x})] \approx \sum_{\{\sigma_I\}} \int_{<} D[\varphi(x)]
  \ e^{-S[\Phi^{\{\sigma_I\}}({\bf x}) + \boldvarphi(x),\delta t({\bf x})]}\quad.
\label{eq:2.8}
\end{equation}
Here $\int_{<}$ indicates that the integration is restricted
to small fluctuations $\boldvarphi$ only.

We now carry out the sum over the saddle point configurations.
The residual interaction between the islands will lead to slight
deviations of the saddle point function from the ideal one
given in (\ref{eq:2.7}). This is taken into
account by replacing the sum over the saddle points
by an integral over a probability distribution
\begin{equation}
P[\Phi] \sim e^{- \frac 1 T \int dx {\cal{L}}^{\rm sp}(\Phi)}~.
\end{equation}
The temperature factor in the
exponent reflects the fact that the saddle points are classical
(static) degrees of freedom \cite{temperature}.
Expanding in powers of the fluctuations, we obtain the following
effective action for the fluctuations $\boldvarphi$
(still for a particular disorder realization)
\begin{eqnarray}
S_{\rm eff} - S^{\rm SP}  &=& S_{\rm G}[\boldvarphi] +S_{\rm int}[\boldvarphi] +
                            S_{\rm cubic}[\boldvarphi] \nonumber \\
                         &+& T \bar{w} \int dx dy \, C(x,y) \sum_{i,j} \varphi_i^2(x)
                         \varphi_j^2(y) \nonumber\\
                         &+& {\rm higher~ order~ terms}~.
\label{eq:2.9}
\end{eqnarray}
The correlation function $C(x,y)$ measures, up to a constant factor determined
by the precise form of $\cal{L}$, whether ${\bf x}$ and ${\bf y}$ belong
to the same island, and $\bar{w} = [(2+4/p) u + 6 \lambda/p]$
is a positive constant.
The $\bar{w}$ term is produced by the interaction
of the fluctuations with the rare regions. It is our approximation of the
effect of these non-perturbative degrees of freedom.
Terms of higher than fourth order in $\boldvarphi$ also
arise, but they are renormalization group irrelevant at both the Gaussian
and the nontrivial fixed points of the conventional theory (see below).

Having identified the effects of the rare regions
we now use the replica trick \cite{replica} to
perform the quenched disorder
average over $\delta t({\bf x})$ which implies an average over position and
size of the rare regions.
The resulting effective action reads
\begin{eqnarray}
&S&_{\rm eff}[\boldvarphi^\alpha(x)]=\
\nonumber\\
&=&
  {1\over 2} \sum_\alpha\,  \sum_{i}\int dx\, dy\,\Gamma_0(x-y)\,
  \boldvarphi_{i}^\alpha(x) \boldvarphi_{i}^\alpha(y)\
\nonumber\\
&+& u\sum_\alpha \sum_{i,j} \int d{\bf x}\, d\tau
\left( {\boldvarphi}_{i}^\alpha({\bf x},\tau) \right)^2
\left( {\boldvarphi}_{j}^\alpha({\bf x},\tau) \right)^2
\nonumber\\
&+& \lambda\sum_\alpha\, \sum_i \int d{\bf x}\, d\tau
\left({\boldvarphi}_{i}^\alpha({\bf x},\tau)\right)^4\
\nonumber\\
&-& \Delta \sum_{\alpha,\beta} \sum_{i,j} \int d{\bf x}\, d\tau d\tau^{\prime}
\left( {\boldvarphi}_{i}^\alpha({\bf x},\tau) \right)^2
\left( {\boldvarphi}_{j}^\beta({\bf x},\tau^{\prime}) \right)^2
\nonumber\\
&-& T\bar{w} \sum_{\alpha,\beta} \sum_{i,j} \int d{\bf x}\, d\tau d\tau^{\prime}
\left( {\boldvarphi}_{i}^\alpha({\bf x},\tau) \right)^2
\left( {\boldvarphi}_{j}^\alpha({\bf x},\tau^{\prime}) \right)^2
\nonumber\\
\label{eq:2.10}
\end{eqnarray}
Here the first four terms are identical to the result of the conventional
treatment. The 5th term has the form of static, annealed disorder and
represents the interaction of the fluctuations with the rare regions in the
Griffiths phase. For more details of this derivation
see Ref.\, \onlinecite{us_rr}.

\section{Renormalization Group Analysis}
\label{sec:III}
\subsection{Flow equations}
\label{subsec:III.A}

We first consider the effective action (\ref{eq:2.10}) at tree level. As usual,
let us define the scale dimension of a length $L$ to be $[L]=-1$, and that of
imaginary time $\tau$ to be $[\tau]=-z$ with $z$ being the dynamical critical exponent.
We first analyze the Gaussian fixed point. From the Gaussian part of the action
(\ref{eq:2.10}) we see that $\omega_n$ scales as $q^2$, implying that $z=2$. The
scale dimension of the field is $[\boldvarphi]=d/2$.
Power counting for the interaction and disorder terms of the action gives the
scale dimensions of $u, \lambda, \Delta$ and $\bar{w}$ as
$[u]=[\lambda]=[\bar{w}]=2-d$ and $[\Delta]=4-d$. Here we have used
the fact that in Matsubara formalism the temperature scales like a frequency, $[T]=z$.
Consequently, $u, \lambda$ and $\bar{w}$ are irrelevant for $d>2$,
while $\Delta$ is irrelevant only for $d>4$.
This implies that in the physical dimension $d=3$ the Gaussian fixed point is
unstable,
and we must carry out a loop expansion of the effective action (\ref{eq:2.10})
close to $d=4$. All terms of higher order in $\boldvarphi$
that arise in addition to those given in (\ref{eq:2.10}) have negative
scale dimensions at and close to $d=4$. Thus, they are irrelevant
by power counting with respect to both the Gaussian and the conventional
non-trivial fixed points.

As in the conventional theory \cite{Yamazaki,bk_afm,dorg} we carry out
the perturbation theory in $d=4-\epsilon$ spatial dimensions and
$\epsilon_{\tau}$ time dimensions. In this way the perturbation
expansion becomes a double expansion in terms of $\epsilon$ and
$\epsilon_{\tau}$.
The renormalization group flow equations are obtained by
performing a frequency momentum shell RG procedure.\cite{Hertz} To one-loop
order, we obtain the following flow equations,
\begin{mathletters}
\label{eqs:3.6}
\begin{eqnarray}
\frac{du}{dl} &=& \tilde{\epsilon}u
                 - 4(p+8)u^2 + 48 u \Delta -24 u \lambda,
\label{eq:3.6a}
\\
\frac{d\lambda}{dl} &=& \tilde{\epsilon} \lambda
                - 36 \lambda^2 + 48 \lambda \Delta -48 u \lambda,
\label{eq:3.6b}
\\
\frac{d\Delta}{dl} &=& \epsilon \Delta + 32 \Delta^2 -
           8(p+2) u \Delta + 8  p  \Delta \bar{w}-24 \Delta \lambda,
\label{eq:3.6c}
\\
\frac{d\bar{w}}{dl} &=& \tilde{\epsilon} \bar{w}
           + 4 p \bar{w}^{2} - 8(p+2) u \bar{w}\ + 48 \Delta
             \bar{w} - 24 \lambda \bar{w}.
\label{eq:3.6d}
\end{eqnarray}
\end{mathletters}
Here we have defined $\tilde{\epsilon}=\epsilon - 2 \epsilon_\tau$. Of course,
also the distance $t$ from the critical point will be renormalized. However,
we only consider the flow on the critical surface $t=0$ since we are interested
in the stability of the critical fixed points. Note that
the coefficient of the rare region term $\bar{w}$ only couples to $\Delta$.
The flow of $u$ and $\lambda$ is only indirectly influenced by the rare regions
(via a modification of the flow of $\Delta$). This will be important later on.

\subsection{Fixed points and their stability}
\label{subsec:III.B}
The flow equations (\ref{eqs:3.6}) possess sixteen fixed points.
Their fixed point values are given in Table \ref{tab:1},
\begin{table*}[t]
\renewcommand{\arraystretch}{1.3}
\begin{tabular}{c|cccc}
\multicolumn{1}{c|}{No.}
      & \multicolumn{4}{c}{FP values}                               \\
      & $u^*$
      & $\lambda^*$
      & $\Delta^*$
      & $\bar{w}^*$                                                         \\
\hline
1     & 0
      & 0
      & 0
      & 0                                                            \\
2     & $\tilde{\epsilon}/4(p+8)$
      & 0
      & 0
      & 0                                                            \\
3     & 0
      & $\tilde{\epsilon}/36$
      & 0
      & 0                                                            \\
4     & $\tilde{\epsilon}/12p$
      & $\tilde{\epsilon}(p-4)/ 36p$
      & 0
      & 0                                                            \\
\hline
5     & 0
      & 0
      & $-\epsilon/32$
      & 0                                                            \\
6     & $(3\epsilon - 2 \tilde{\epsilon})/16(p-1)$
      & 0
      & $[(p+8)\epsilon -2(p+2)\tilde{\epsilon}]/64(p-1)$
      & 0                                                            \\
7     & 0
      & ${\cal{O}}(\epsilon^{1/2})$
      & ${\cal{O}}(\epsilon^{1/2})$
      & 0                                                            \\
8     & $(3\epsilon-2\tilde{\epsilon})/24(p-2)$
      & $[(3\epsilon-2\tilde{\epsilon})(p-4)]/72(p-2)$
      & $[3p\epsilon -4(p-1)\tilde{\epsilon}]/96(p-2)$
      & 0                                                            \\
\hline
9     & 0
      & 0
      & 0
      & $-\tilde{\epsilon}/4p$                                         \\
10    & $\tilde{\epsilon}/4(p+8)$
      & 0
      & 0
      & $[(p-4) \tilde{\epsilon}]/4p(p+8)$                                                    \\
11    & 0
      & $\tilde{\epsilon}/36$
      & 0
      & $-\tilde{\epsilon}/12p$                                                            \\
12    & $\tilde{\epsilon}/12p$
      & $[(p-4)\tilde{\epsilon}]/ 36p$
      & 0
      & $(p-4)\tilde{\epsilon}/12p^2$                                                            \\
\hline
13    & 0
      & 0
      & $(\epsilon-2\tilde{\epsilon})/64$
      & $(2\tilde{\epsilon}-3\epsilon)/16p$                                                            \\
14    & $(3\epsilon-2\tilde{\epsilon})/8(10-p)$
      & 0
      & $[(p+8)\epsilon-12\tilde{\epsilon}]/32(10-p)$
      & $[(3\epsilon-2\tilde{\epsilon})(p-4)]/ 8p(10-p)$                      \\
15    & 0
      & $(3\epsilon-2\tilde{\epsilon})/72$
      & $(9\epsilon-12\tilde{\epsilon})/288$
      & $-3(3\epsilon-2\tilde{\epsilon})/72p$                                                            \\
16    & $(3\epsilon-2\tilde{\epsilon})/48$
      & $(3\epsilon-2\tilde{\epsilon})(p-4)/144$
      & $(3p\epsilon-2(p+2)\tilde{\epsilon})/192$
      & $(3\epsilon-2\tilde{\epsilon})(p-4)/48p$                                     \\
\end{tabular}
\vspace{2mm}
\caption{Fixed points of the flow equations,
 $p$ is the number of order parameter components. }
\label{tab:1}
\end{table*}
the eigenvalues of the corresponding linearized renormalization group
transformations are listed in Table \ref{tab:2}.
\begin{table*}[t]
\renewcommand{\arraystretch}{1.3}
\begin{tabular}{c|cccc}
\multicolumn{1}{c|}{No.}
      & \multicolumn{4}{c}{eigenvalues}                               \\
      & $e_1$
      & $e_2$
      & $e_3$
      & $e_4$                                                         \\
\hline
1     & $\tilde{\epsilon}$
      & $\tilde{\epsilon}$
      & $\epsilon$
      & $\tilde{\epsilon}$                                                            \\
2     & $-\tilde{\epsilon}$
      & $(p-4)\tilde{\epsilon}/(p+8)$
      & $\epsilon-2(p+2)\tilde{\epsilon}/(p+8)$
      & $-(p-4)\tilde{\epsilon}/(p+8)$                                     \\
3     & $\tilde{\epsilon}/3$
      & $-\tilde{\epsilon}$
      & $\epsilon-2\tilde{\epsilon}/3$
      & $\tilde{\epsilon}/3$                                                            \\
4     & $-\tilde{\epsilon}$
      & $-\tilde{\epsilon}(p-4)/ 3p$
      & $\epsilon-4\tilde{\epsilon}(p-1)/3p$
      & $-\tilde{\epsilon}(p-4)/ 3p$                                   \\
\hline
5     & \multicolumn{4}{l}{eigenvalues not calculated since FP is unphysical}  \\
6     & $\frac{-A + \sqrt {A^2 - B}}{p - 1}$
      & $\frac{-A - \sqrt {A^2 - B}}{p - 1}$
      & $(p-4)(3\epsilon -2\tilde{\epsilon})/4(p-1)$
      & $-(p-4)(3\epsilon -2\tilde{\epsilon})/4(p-1)$                \\
7     & ${\cal{O}}(\epsilon^{1/2})$
      & ${\cal{O}}(\epsilon^{1/2})$
      & ${\cal{O}}(\epsilon^{1/2})$
      & ${\cal{O}}(\epsilon^{1/2})$                                  \\
8     & $\frac{-E + \sqrt {E^2 - F}}{12(p - 2)}$
      & $\frac{-E - \sqrt {E^2 - F}}{12(p - 2)}$
      & $-(3\epsilon-2\tilde{\epsilon})(p-4)/6(p-2)$
      & $-(3\epsilon-2\tilde{\epsilon})(p-4)/6(p-2)$                   \\
\hline
9     & \multicolumn{4}{l}{eigenvalues not calculated since FP is unphysical}  \\
10    & $-\tilde{\epsilon}$
      & $(p-4)\tilde{\epsilon}/(p+8)$
      & $\epsilon-12\tilde{\epsilon}/(p+8)$
      & $(p-4)\tilde{\epsilon}/(p+8)$              \\
11    & \multicolumn{4}{l}{eigenvalues not calculated since FP is unphysical}  \\
12    & $-\tilde{\epsilon}$
      & $-\tilde{\epsilon}(p-4)/ 3p$
      & $\epsilon-2\tilde{\epsilon}(p+2)/3p$
      & $\tilde{\epsilon}(p-4)/ 3p$                                       \\
\hline
13    & \multicolumn{4}{l}{eigenvalues not calculated since FP is unphysical}  \\
14    & $\frac{-C + \sqrt {C^2 - D}}{4\,(10- p)}$
      & $\frac{-C - \sqrt {C^2 - D}}{4\,(10- p)} $
      & $(p - 4)(3\epsilon -2 \tilde{\epsilon})/{2(10 - p)}$
      & $(p - 4)(3\epsilon -2 \tilde{\epsilon})/{2(10 - p)}$ \\
15    & \multicolumn{4}{l}{eigenvalues not calculated since FP is unphysical}  \\
16    & $\frac{-G + \sqrt {G^2 - H}}{24}$
      & $\frac{-G - \sqrt {G^2 - H}}{24} $
      & $(3\epsilon-2\tilde{\epsilon})(p-4)/12$
      & $-(3\epsilon-2\tilde{\epsilon})(p-4)/12$                                     \\
\end{tabular}
\vspace{2mm}
\caption{Eigenvalues of the corresponding linearized RG transformation.
 $p$ is the number of order parameter components.
 $A$, $B$, $C$, and $D$ are defined as
 $A = (p+8) \epsilon - 2(p-4) \tilde{\epsilon}$,
 $B = 16(p-1)\,(3 \epsilon-2\tilde{\epsilon})\,[(p+8) \epsilon - 2(p+2) \tilde{\epsilon}]$,
 $C = (p+8)\epsilon -2(p-4)\tilde{\epsilon}$,
 $D = 8(10 - p)\,(3\epsilon  -2\tilde{\epsilon})\,
   [8\epsilon -12 \tilde{\epsilon}+p\epsilon]$.
 Analogously,
 $E=3p\epsilon + 2(p-4)\tilde{\epsilon}$,
 $F=24(p-2)\,(3\epsilon-2\tilde{\epsilon})\,[4\tilde{\epsilon} + 3p\epsilon -4p\tilde{\epsilon}]$,
 $G=8\tilde{\epsilon}+3p\epsilon-2p\tilde{\epsilon}$,
 $H=48(3\epsilon-2\tilde{\epsilon})[-4\tilde{\epsilon} +3 p\epsilon -2p\tilde{\epsilon}]$.
}
\label{tab:2}
\end{table*}
For eight of the sixteen fixed points (Nos. 1--8 in Table \ref{tab:1})
the fixed point value of the rare region term is $\bar{w}^* = 0$.
These fixed points have already been studied in
Ref.\, \onlinecite{Yamazaki} using the conventional approach.
In the following, we concentrate on the case $\epsilon >0$ and
$\tilde{\epsilon}=\epsilon-2\epsilon_\tau <0$ relevant for the itinerant
quantum antiferromagnet.

We first consider the dirty Heisenberg fixed point (No. 6) and
the dirty cubic fixed point (No. 8). These are the stable fixed points
of the conventional theory for the cases of $p<4$ and $p>4$, respectively.
Analyzing the stability matrix for the dirty Heisenberg fixed point
shows that it is unstable since the eigenvalue
$e_4$ is positive for $p<4$. In contrast, the dirty cubic fixed
point remains stable for $p>4$ since all eigenvalues of the stability matrix
are negative. Thus we conclude that the rare regions destroy the
conventional dirty Heisenberg critical behavior for $p<4$ while they do not
influence the conventional dirty cubic critical behavior for $p>4$.

We now turn to the new fixed points with $\bar{w}^* \neq 0$ (Nos. 9 -- 16
in Table \ref{tab:1}).  Fixed points 9, 11, 13 and 15 are unphysical
because their fixed point values $\bar{w}^*$ are negative. Since the bare
$\bar{w}$ is positive and according to eq. (\ref{eq:3.6d})
the flow cannot cross the $(\bar{w}=0)$-plane
these fixed points can never be reached. Depending on the number $p$ of
order parameter components the remaining fixed points
(Nos. 10, 12, 14, and 16) are either also unphysical, or they are unstable.
Consequently, for $p<4$ and to one-loop order there is no stable fixed point.
Renormalization group trajectories which in the conventional theory
would go to the dirty Heisenberg fixed point show runaway
flow to large disorder strength. This runaway flow could
either indicate a unconventional phase transition, e.g. an infinite
disorder critical point as in the one-dimensional random Ising
model \cite{Fisher} or a percolative rather than a homogeneous transition
or even a destruction of the phase transition. Within the present
approach we cannot be decide between these alternatives.

The influence of the rare regions on the stability of the fixed points
in our model is similar to that in the isotropic case \cite{us_rr}.
For $p<4$ the conventional fixed point is destroyed in both models.
For $p>4$ the conventional fixed point is stable. In our model this is
the dirty cubic fixed point while in the isotropic case this stable fixed
point is the dirty Heisenberg fixed point.

\subsection{The fluctuation-driven first-order transition}
\label{subsec:III.C}
In addition to the continuous phase transitions associated with
the critical points discussed above there is also the possibility
for a first-order transition in the model considered here.
Let us first discuss the mechanism for a clean system and discuss
the effects of disorder and rare regions later.

According to a mean-field stability analysis of the effective action
(\ref{eq:2.10}) with $\Delta=\bar{w}=0$ the inequalities
$u+\lambda>0$ (for $u>0$) and $u+\lambda/p >0$ (for $u<0$) have to be
fulfilled for the theory to be stable. Now consider a bare theory with
$u<0, \lambda>0$ or $u>0, \lambda<0$ but still fulfilling the
above stability conditions.  In these cases the flow equations
(\ref{eqs:3.6}) can lead the renormalization group trajectories
to the mean-field unstable region. This indicates a fluctuation-driven
first-order transition \cite{Rudnick,Amit}.
It was later shown \cite{Yamazaki2,Cardy}
that the fluctuation-driven first-order in this
model survives the presence of quenched disorder, at least within the
conventional theory. Let us now consider the influence of
the rare regions. As already mentioned, the rare region coefficient $\bar{w}$
does not couple into the flow equations for $u$ and $\lambda$ but only into
the flow equation for $\Delta$.  Thus a renormalization group trajectory
going to the mean-field unstable region within the conventional theory will
generically also do so in the presence of rare regions, the only modification
being a different disorder value at the stability boundary.

Therefore, we conclude that the fluctuation-driven first-order
transition also occurs when taking the rare regions into account.
However, since the rare regions modify the flow of the disorder
strength $\Delta$, the boundaries of the first-order region may
change.

\section{Summary and conclusions}
\label{sec:IV}
We have investigated the influence of rare regions on the
quantum phase transition of a disordered itinerant antiferromagnet
with cubic anisotropy. The local magnetic moments forming on
the rare regions in the Griffiths phase generate a new term
in the order parameter field theory which has the form of static annealed
disorder. We have found that for order parameter dimension
$p>4$ this new term does not change the critical behavior,
which is characterized by the dirty cubic fixed point.
In contrast, for $p<4$ the rare region term renders the conventional critical
fixed point unstable. The renormalization group trajectories
show runaway flow to large disorder. Within our approach
which is essentially perturbative, even though it includes
some non-perturbative degrees of freedom (the local moments)
we cannot determine the ultimate fate of the transition.
It could be an unconventional phase transition, e.g. an infinite
disorder critical point or a percolative rather than a homogeneous
transition or even the destruction of the phase transition.
We have also found that the fluctuation-driven first-order
transition occurring in this model remains qualitatively
unchanged by the rare regions, while the precise position of
the first-order region in parameter space will change.

The authors acknowledge helpful discussions with D.\ Belitz, J.\ Cardy, and
T.R.\ Kirkpatrick. R.N. thanks the hospitality of TU Chemnitz during two vistits
where part of the research was performed.
This work was supported in part by the DFG under grant nos.
SFB393/C2 and Vo659/2, by the NSF under grant no. DMR-98-70597, and by
EPSRC under grant no. GR/M 04426.

\end{document}